\documentclass[aps,prx,twocolumn,floatfix,superscriptaddress,groupaddress, longbibliography, 10pt]{revtex4-1} 
\usepackage{graphicx}  
\usepackage{color}
\usepackage{dcolumn}   
\usepackage{bm}        
\usepackage{amssymb}   
\usepackage{amsmath,textcomp}
\usepackage{float}
\usepackage[hidelinks, bookmarks=false]{hyperref}

\usepackage{natbib}

\begin{document}

\title{Detecting Acoustic Blackbody Radiation with an Optomechanical Antenna}

	\author{Robinjeet Singh}
	\email{rsingh17@umd.edu}
      \affiliation{Joint Quantum Institute, University of Maryland, College Park 20742 USA}
			  			\affiliation{National Institute of Standards and Technology, Gaithersburg, Maryland 20899, USA}
					\author{Thomas P. Purdy}
					\email{tpp9@pitt.edu}
				\affiliation{National Institute of Standards and Technology, Gaithersburg, Maryland 20899, USA}
			\affiliation{Department of Physics $\&$ Astronomy, University of Pittsburgh, Pittsburgh 15260, USA}
	
	\date{\today}

\date{\today}

\begin{abstract}
Nanomechanical  systems are generally embedded in a macroscopic environment where the sources of thermal noise are difficult to pinpoint. We engineer a silicon nitride membrane optomechanical resonator such that its thermal noise is acoustically driven by a spatially well-defined remote macroscopic bath. This bath acts as an acoustic blackbody emitting and absorbing acoustic radiation through the silicon substrate. Our optomechanical system acts as a sensitive detector for the blackbody temperature and for photoacoustic imaging. We demonstrate that the nanomechanical mode temperature is governed by the blackbody temperature and not by the local material temperature of the resonator. Our work presents a route to mitigate self-heating effects in optomechanical thermometry and other quantum optomechanics experiments, as well as acoustic communication in quantum information.

\end{abstract}

\maketitle

	Understanding the origins of dissipation and noise in nanomechanical resonators is vital to a wide range of applications including ultra-sensitive force transduction and the production of long-lived quantum states of macroscopic objects. With much dedicated recent work, a variety of mechanical resonator loss channels have been identified and mitigated, leading to  unprecedentedly low dissipation in room temperature nanoscale systems~\cite{Ghadimi2018,Tsaturyan2017}. Quite generally, these loss channels dissipate energy into baths that also inject random thermal noise back into the nanomechanical resonator, a consequence of a fluctuation-dissipation theorem~\cite{Callen1951}. In the absence of additional forces, the resonator comes to thermal equilibrium with its coupled bath, and the scale of its Brownian motion is dictated by the bath temperature. The bath need not be physically localized in the device itself, as with clamping or anchor loss where acoustic energy is radiated into a supporting substrate~\cite{WilsonRae2008}, viscous damping by a surrounding fluid, or laser cooling of mechanical motion~\cite{Peterson2016}. In these cases, the mechanical mode temperature is not set by the local temperature of the material from which the nanoscale resonator is made, but by the temperature of the external macroscopic bath. Such externally damped systems usefully show immunity to self-heating effects from power absorbed by optical or electrical probing and offer a route to remote temperature sensing. Here, we investigate the dissipation and thermal noise in an optically detected, silicon nitride (SiN) membrane mechanical resonator whose mechanical loss is dominated by external acoustic radiation, exploring a technique capable of drastically reducing systematic uncertainty in optomechanical noise thermometry~\cite{Purdy2017}, improving low-temperature thermalization in quantum optomechanical systems~\cite{Patel2017,Riedinger2018}, and presenting a path toward acoustic communication of quantum information~\cite{KCBalram2016, Fang2016}.

	\begin{figure}[H]
\begin{center}
\includegraphics[width=1\columnwidth]{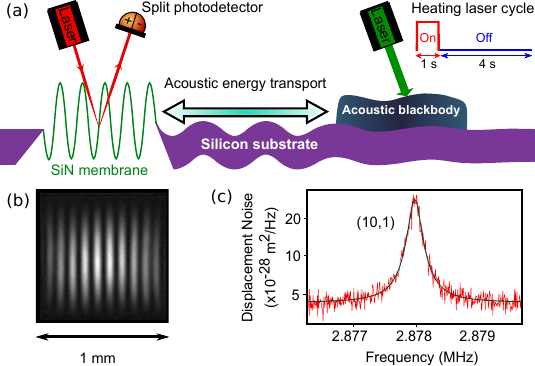}
\caption{Acoustic blackbody thermometry. (a) A silicon substrate acts as an acoustically transparent medium coupling a SiN membrane mechanical resonator to a remote acoustic blackbody. Thermal acoustic radiation emitted into the substrate by the blackbody drives Brownian motion of the resonator.  The resonator motion is read out with a simple optical-lever detection scheme. (b) Dark-field image~\cite{Chakram2014} of the (10,1) resonator mode. (c) Displacement noise spectrum of a thermally occupied (10,1) mode. }
\label{Schematic}
\end{center}
\end{figure}

 Our experiment (Fig.~\ref{Schematic}), consists of an optically probed membrane mechanical resonator that exchanges energy with a remote bath made of mechanically lossy material via acoustic radiation through their common substrate. The lossy material acts as a broadband absorber and emitter of acoustic energy -- an ``acoustic blackbody"~\cite{Mellen1952}. In essence, when measuring the Brownian motion of the resonator, we perform an acoustic analog of non-contact infrared thermometry or thermal imaging. In the case of infrared thermometry, electromagnetic blackbody radiation is transmitted through a transparent medium to a remote detector.  In our case, the acoustic blackbody, subject to large, random thermal force fluctuations, emits acoustic radiation into the low-mechanical-loss silicon substrate, which is detected as motion of the nanomechanical resonator.

\begin{figure}
\begin{center}

\includegraphics[width=1\columnwidth]{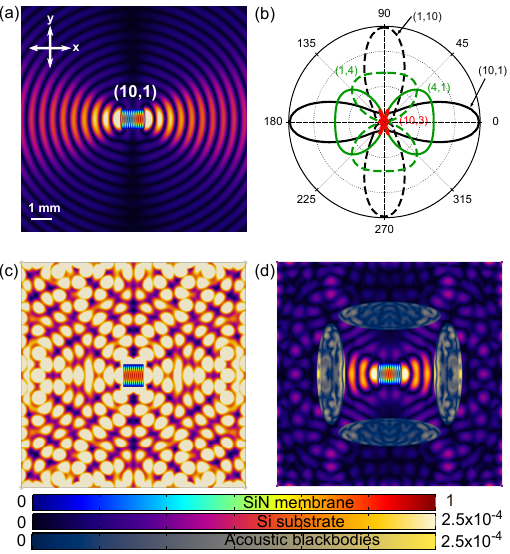}
\caption{Finite element simulations of the acoustic radiation pattern of the modes of a 1 mm square by 200 nm thick SiN membrane resonator on a 200 $\mu\text{m}$ thick silicon substrate. (a) Radiation pattern of (10,1) mode into a substrate of infinite transverse extent.  (b) Far field radiation pattern  of several modes normalized to their energy damping rate.  Modes of the form ($i$,1) show high directionality for $i \gg 1$, whereas, {\it e.g.}, the (10,3) mode displays multiple small lobes of radiation.  (c) Radiation pattern of mode (10,1) into a finite extent, square silicon substrate (10 mm side length). (d) Simulated model of the device engineered for the experiment. Four oblong acoustic blackbodies (modeled with mechanical loss tangent of 0.1), offer high absorption in a single pass, eliminating interference due to edge reflections and hence maintaining the directionality of the acoustic wave.  Simulation geometry is chosen to match the experimental device parameters. Color schemes represent relative displacement of the membrane, silicon substrate, and acoustic blackbodies. Color bars are same for a, c, and d.}
\label{FEA}
\end{center}
\end{figure}

	We focus on the out-of-plane drumhead modes of a square, high-tensile-stress SiN membrane suspended from a silicon substrate, with the mode indices $i$ and $j$ representing the number of antinodes along the horizontal, $x$-direction, and vertical, $y$-direction, respectively (see Fig.~\ref{FEA}).  These high quality factor (Q) SiN resonators~\cite{Thompson2008,Verbridge2008} have become important building blocks of optomechanical systems in the quantum regime~\cite{Purdy2013} and have recently achieved record-low dissipation at room temperature by mitigating multiple loss channels~\cite{Ghadimi2018,Tsaturyan2017, Norte2016,Chakram2014, Yu2014, Tsaturyan2014, Rieger2014}.  The mechanical resonance frequencies of the drumhead modes are increased by orders of magnitude due to tensile stress, which dilutes the internal, bending losses, compared to those of stress-free films~\cite{Southworth2009}. Then, for many membrane mechanical modes, especially those with certain symmetry~\cite{WilsonRae2011}, radiative damping becomes the dominant dissipation channel. 
	
	The temperature of a mechanical mode $T^{(i,j)}$, ascertained from its average Brownian motion, $\langle |x^{(i,j)}|^2\rangle$, is set by the couplings to its dissipation channels,
	
	\begin{eqnarray}
T^{(i,j)} \equiv \frac{m_\text{eff} (\omega_m^{(i,j)})^2}{k_{B}} \langle |x^{(i,j)}|^2\rangle = \sum_k \frac{\Gamma^{(i,j)}_{k}}{\Gamma^{(i,j)}} T_k
\label{e1}
\end{eqnarray}

where $\omega_m^{(i,j)}$, $m_\text{eff}$, and $\Gamma^{(i,j)}$ are the mechanical frequency, effective mass, total mechanical damping rate of mode $(i,j)$, respectively, $\Gamma^{(i,j)}_k$ is the energy damping rate of the mode $(i,j)$ into the $k^{th}$ dissipation channel, such that $\sum_k \Gamma^{(i,j)}_k=\Gamma^{(i,j)}$, and $T_k$ is the bath temperature of the $k^{th}$ dissipation channel.  Typically, several dissipation channels may contribute, such as surface effects~\cite{Villanueva2014}, thermoelastic damping~\cite{Lifshitz2000}, or acoustic radiation into various spatial directions~\cite{WilsonRae2008}.  Baths associated with each channel may be at different temperatures. For example, electron and phonon temperatures may differ in cryogenic systems, or large thermal gradients may be generated by localized heating from active elements or absorbed light. In general, a nanomechanical resonator will always display Brownian motion, regardless of what baths it is coupled to, making it difficult to disentangle the origin of thermal noise.  So we have performed several experiments and simulations to  demonstrate that indeed the mechanical noise measured in our membrane resonator originates from the remote acoustic blackbody that we have engineered.	 

Our basic strategy is to create two membrane mechanical modes that are radiatively coupled to distinct and spatially separate acoustic blackbodies. This configuration enables us to independently control the temperature of a given mechanical mode by  varying the temperature of the blackbody to which it is coupled. To design a device with the above properties, we must first understand the acoustic radiation pattern of the membrane modes into the substrate.  We start by modeling a finite thickness, but infinite transverse extent substrate with a finite element simulation.  For our megahertz frequency devices the typical wavelength in the substrate is much longer than the substrate thickness, so the acoustic radiation is dominated by flexural excitations, specifically the first-order, symmetric Lamb wave mode (Fig.~\ref{Schematic}(a)).  We note that for higher frequencies or thicker substrates, the dominant form of radiation would become Rayleigh, surface acoustic waves.  The transverse radiation pattern is distinct for each mode, with several examples shown in Fig.~\ref{FEA}(b).  We also note that each membrane mode acts as a reciprocal antenna~\cite{Strutt1871}, being efficiently excited by the same acoustic radiation pattern that it emits into the substrate.

  We find that for modes of the form $(i,1)$, with $i\gg 1$, the energy radiated becomes very strong and the far-field radiation pattern becomes highly directional, concentrated in the $x$-direction, while for modes $(1,j)$, radiation is concentrated in the $y$-direction.  Whereas, other modes with more nodal lines in both the horizontal and vertical directions generally radiate less energy and exhibit several radiation lobes in various directions, for example $(10,3)$ mode in Fig~\ref{FEA}(b).  Intuitively, this can be understood using the phonon tunneling approach~\cite{WilsonRae2008,WilsonRae2011}, where acoustic waves emitted out of phase from different segments of the membrane perimeter interfere destructively to limit the far field radiation pattern in particular directions.  For a substrate of finite transverse extent and relatively low acoustic loss, acoustic radiation cannot easily escape the system. Our simulations show that the radiation tends to create complex interference patterns as it reflects off the edges of the substrate  (see  Fig.~\ref{FEA}(c)). The exact shape of these patterns varies critically with the substrate dimensions, often leading to unpredictable and irreproducible Q due to dissipation at the device mounting points~\cite{Wilson2009}.

		Guided by these simulations, we work to engineer our system so that, in particular, the $(10,1)$ and $(1,10)$ membrane modes have well-defined acoustic radiation channels that dominate the dissipation of these modes.   Our device consists of a 1~mm square by 200~nm thick stoichiometric Si$_3$N$_4$ membrane suspended from a 10~mm square by a 200~$\mu$m thick silicon substrate. The device is weakly mounted at its corners to an aluminum heat sink minimizing dissipation at the mounting points.  To readout mechanical motion of the membrane, we measure the deflection of a laser beam reflecting from the membrane.  To define our baths, we deposit acoustic blackbodies, formed from a mechanically lossy epoxy composite, on the substrate in the path of the radiation from a particular resonator mode. The baths consist of two oblong blackbodies along the $x$-direction, bath X, and two along the $y$-direction, bath Y.  The size and thickness of the epoxy is chosen so that a large fraction of radiation from the membrane is absorbed in a single pass as illustrated in Fig.~\ref{FEA} (d). Because the blackbody has high absorptivity, we also expect it to have high emissivity~\cite{Kirchhoff1860} for the relevant acoustic modes.  For this device, the mechanical Q's of the $(10,1)$ and $(1,10)$ modes are about 14,000, while for similar devices without the baths, the Q's for these modes are typically 2-3 orders of magnitude larger.  This drop in Q is a strong evidence that the mechanical loss for these modes is dominated by acoustic radiation into the baths.  Additionally, we find that modes predicted to exhibit low acoustic radiation loss, even with the acoustic blackbodies present have Q's on the order of $5\times10^5$.
		
\begin{figure}[H]
\begin{center}
\includegraphics[width=1\columnwidth]{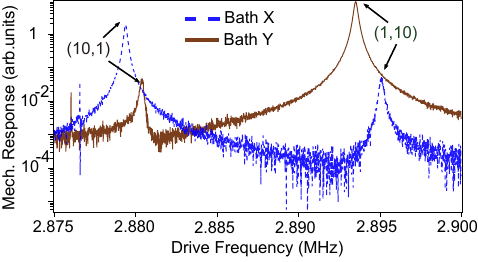}
\caption{Photoacoustically driven frequency response of two membrane modes to periodic excitation of bath X (blue) and bath Y (brown), showing the directional sensitivity of our detection scheme. The mode frequencies shift between the two traces due to thermally induced stress (see main text).}
 \label{Spectrogram}
\end{center}
\end{figure}

		Before measuring the thermal noise of the device, we investigate the coupling of the acoustic blackbody to membrane modes via photoacoustic imaging~\cite{Yao2014,gusev1993,wang2017}.  We coherently drive motion of one of the acoustic blackbodies via thermal expansion induced by the absorption of approximately 100~mW of sinusoidally amplitude-modulated laser light, producing an AC heat load.  Acoustic energy from this drive is then radiated into the substrate and excites the membrane modes well above their thermal noise floors.  We sweep the frequency of the photoacoustic excitation and measure the response of the membrane as shown in Fig.~\ref{Spectrogram}. Consistent with our expectations, we observe that driving bath X strongly excites only the $(10,1)$ mode, while driving bath Y, strongly excites only the $(1,10)$ mode, showing that acoustic wave transduction in our device is highly directional. We expect this two-order-of-magnitude separation in response between the two mechanical modes to persist for thermal force noise exciting the baths. 
		

		We note that our photoacoustic testing already represents a very sensitive ultrasonic receiver.  Previously, traditional piezoelectric ultrasonic transducers have been used in experiments focused on imaging applications of thermal acoustic radiation~\cite{Weaver2001, Mansfeld2009}.  Such transducers are limited by intrinsic thermal and electronic amplifier noise and do not reach the regime where thermal radiation largely dominates the signal.  Optomechanical transducers~\cite{Bowen2019} with low intrinsic dissipation and quantum-limited optical readout are able to reach this regime.  In our device, the resonant thermal noise is well above the shot-noise-limited optical readout noise floor (Fig.~\ref{Schematic}(c)).  Based on the ratio of intrinsic Q to radiatively damped Q and the measurements described below, the thermal acoustic noise dominates over intrinsic thermal noise by 2-3 orders of magnitude in our system. Using our finite element model, we estimate the intrinsic thermal noise limited sensitivity of our device to the amplitude of resonant excitations in the substrate to be on the scale of $\mathrm{attometers}/\sqrt{\mathrm{Hz}}$.

\begin{figure}
\begin{center}

\includegraphics[width=1\columnwidth]{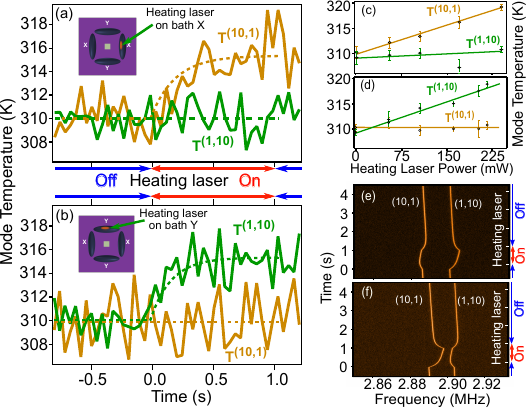}

\caption{Thermometry results. (a) and (b) Brownian motion temperature profiles, derived from the area under the spectrogram, of (10,1) mode (orange) and (1,10) mode (green), when bath X (a) or bath Y (b) is heated with a 1 s laser pulse. Dotted lines represent the expected temperature rise using $\tau_{BB}$ derived from the mechanical frequency shifts. In each case, the mode strongly coupled to the heated bath shows a rise in temperature, while the uncoupled mode stays at a constant temperature to within the measurement uncertainty. (c) and (d) Temperature rise as a function of heating laser power. Error bars represent the $1 \sigma$ statistical uncertainty from a few thousand experimental trials. Colored lines are linear fits to the data. (e) and (f) Spectrograms of the resonator motion showing the thermally driven noise peaks of mechanical modes, taken when either bath X (e) or bath Y (f) are heated. Mode frequencies shift at two time scales following $\tau_{BB} \approx 100$~ms, and the entire apparatus (several 10's of seconds). Logarithmic color-scale.}

 \label{Temps}
\end{center}
\end{figure}

			We next investigate the ballistic transport of thermal acoustic radiation through the silicon substrate.  We heat one of the blackbodies forming bath X with a 1~s pulse of laser light, which elevates the average temperature of bath X on the order of 10 K. This pulse is not amplitude modulated at the mechanical frequency. Here, acoustic radiation is generated by the thermal motion of the blackbody. The heating saturates with a time constant, $\tau_{BB}\sim 100$~ms, coming to equilibrium on a time scale and with a magnitude consistent with a simple 1D model for heat diffusion through the thickness of the blackbody.

			We create a spectrogram of the membrane motion  by Fourier transforming short time intervals of our signal collected from the optical detection system. Figure \ref{Temps}(e, f) shows the average spectrogram over a few thousand heating laser pulses.  To estimate the mode temperatures of the membrane, we find the spectral area under each mechanical resonance peak for each line of the averaged spectrogram. To calibrate this data we assume the equilibrium temperature of each mode is given by a calibrated resistive thermometer used to stabilize the temperature of the sample mount to 310~K.  This temperature is consistent with one calculated using direct measurements of the mechanical to optical conversion efficiency ascertained by measuring the optical response to a known angular deviation of the probe laser path. The 10~\% level uncertainty in this measurement is dominated by uncertainty in the optical path length and knowledge of the membrane modal mass.

		Fig.~\ref{Temps} shows that when we apply heat to bath X, the $(10,1)$ mode displays a significant rise in temperature, tracking the rise in temperature of bath X, while the $(1,10)$ mode remains at a constant temperature.  The opposite holds when we apply a heat pulse to bath Y, where only the $(1,10)$ mode increases in temperature.  This behavior is conclusive evidence that dissipation in the $(10,1)$ mode is dominated by acoustic radiation into bath X and is uncoupled to bath Y. Thus, we have performed a remote, spatially resolved, acoustically mediated thermometry, where different modes of our membrane act as directional antennas for thermal noise.

It is important to note that the membrane and bath are coupled through the substrate via ballistically transported acoustic energy over millimeter distances, while diffusive heat transport plays very little role.  At the sub-micron scale, many systems show enhanced thermal conductivity due to ballistic transport of phonons~\cite{Balandin2011, Johnson2013, Anufriev2017}. Through a heat transfer model of our system, we estimate that the temperature rise of the membrane material due to heat diffusion is a factor of ten less than the expected temperature rise from acoustic blackbody radiation. We further note that the time scale to heat up the entire apparatus is much longer than the experimental cycle.  Also, the mechanical ringdown time of the resonator mode is much faster than any other time constants of our system. So we can safely assume that the mechanical mode temperature adiabatically follows the temperature evolution of the coupled acoustic blackbody.

	Also evident in the spectrograms of Fig.~\ref{Temps}(e, f) is a shift in mechanical resonance frequencies at the time scale $\tau_{BB}$.  We attribute this shift to the thermal strain in the blackbody material which induces nonuniform stress in the membrane. The $(10,1)$ and $(1,10)$ modes are highly sensitive to variations in the $x$ and $y$ components of the membrane stress tensor, respectively, and relatively insensitive to the orthogonal stress component.  With our device geometry, thermally induced strain from heating bath X reduces the $x$ component of the stress tensor, while increasing the $y$ component.  This produces frequency shifts of opposite sign and similar magnitude for the $(10,1)$ and $(1,10)$ modes. All of the modes also experience a weak downward shift in frequency due to the bulk heating of the entire apparatus on a time scale much larger than the experimental duration. In our analysis of the Brownian motion of the membrane, we take into account these frequency shifts when computing mode temperatures from the measured average amplitude of motion (Eq.~1). 
	
	In other frequency bands, we have observed mode hybridization and avoided crossings in the spectrograms, when pairs of membrane modes are brought into near degeneracy by thermal strain.  We have intentionally chosen to work with the $(10,1)$ and $(1,10)$ modes, which always remain nondegenerate.  The spatial profiles of the hybridized modes change both the modal effective mass and the optical transduction efficiency, making it difficult to accurately convert measured levels of motion into temperature.  Also, in this case, we observe non-Lorentzian lineshapes in the thermal noise spectrum, as seen in other mechanical systems with inhomogeneously distributed mechanical loss~\cite{Schwarz2016, Yamamoto2001}.

In summary, we have developed a nanomechanical system with strong acoustic blackbody coupling to a well-defined macroscopic bath. We have used a single nanomechanical degree of freedom to perform high sensitivity noise thermometry of this remote bath. This principle of separating device from bath ensures that the nanomechanical mode temperature is nearly independent of the SiN membrane material temperature, providing immunity to local self-heating effects of probe light absorbed by the membrane. Self-heating is a dominant source of uncertainty and noise in quantum optomechanical thermometry~\cite{Purdy2017} and other low temperature quantum optomechanics experiments~\cite{Meenehan2014}. The thermal bath engineering technique that we have developed can be readily applied to different size and frequency scale of nanomechanical systems~\cite{Sarabalis2017, Patel2017} to improve the accuracy of optomechanical temperature metrology. Extending our work by replacing the acoustic blackbodies with cavity optomechanical transducers~\cite{Fang2016} capable of both optical detection and actuation, we could carefully map out the directionality of the radiation by measuring the acoustic coupling between transducers. This scheme can provide a building block for quantum acoustic networks where information is coherently transduced between quantum optomechanical systems.

\begin{acknowledgments}

We thank Vladimir Aksyuk, Marcelo Wu, and Alan Migdall for useful discussions.    
\end{acknowledgments}

%


\end{document}